\documentclass[showpacs,twocolumn,eqsecnum,aps,amssymb,nofootinbib,prd]{revtex4}
\usepackage{graphicx, epsfig, times, bm}

\newcommand{\be}{\begin{equation}}
\newcommand{\ee}{\end{equation}}
\newcommand{\bea}{\begin{eqnarray}}
\newcommand{\eea}{\end{eqnarray}}

\newcommand{\PR}{{\it Phys. Rev.\,}}

\newcommand{\PL}{{\it Phys. Lett.\,}}

\begin{document}

\title{Why the Universe Started from a Low Entropy State}

\author{R.~Holman}

\affiliation{Department of Physics, Carnegie Mellon University, Pittsburgh PA 15213, USA}

\email[]{rh4a@andrew.cmu.edu}

\author{L.~Mersini-Houghton}

\affiliation{Department of Physics and Astrononmy, UNC-Chapel Hill, NC, 27599-3255, USA}

\email[]{mersini@physics.unc.edu}

\date{\today}
 
\begin{abstract} 
We show that the inclusion of backreaction of massive long wavelengths imposes dynamical constraints on the allowed phase space of initial conditions for inflation, which results in a superselection rule for the initial conditions. Only high energy inflation is stable against collapse due to the gravitational instability of massive perturbations. We present arguments to the effect that the initial conditions problem {\it cannot} be meaningfully addressed by thermostatistics as far as the gravitational degrees of freedom are concerned. Rather, the choice of the initial conditions for the universe in the phase space and the emergence of an arrow of time have to be treated as a dynamic selection.
\end{abstract}


\maketitle

\section{Introduction}


The landscape picture of string vacua\cite{landscape} has been the driving influence behind a major rethinking of the way that predictions are extracted from a physical theory. One view is that we are necessarily led to something like the anthropic principle\cite{anthland} as the only way that string theory can be made predictive. We must then hope that the physical quantities of interest are amenable to anthropic predictions; an example of this is Weinberg's calculation of the cosmological constant. An interesting discussion of this and other points concerning the landscape can be found in Ref.\cite{nima}.

Another point of view is that string theory might still be as predictive as other theories in physics, but that it requires a much deeper understanding of the initial conditions (IC) for the Universe. In \cite{shortrichlaura,laurareview} we argued that a landscape picture {\em must}, in fact, be expected of any theory of initial conditions. The hope would then be that a superselection rule emerges from the quantum dynamics whereby the Universe would find itself driven to choose a {\em unique} vacuum state which could be computed from the requirement that it is the most probable state the Universe can access starting from ``nothing'', \cite{land1,land2,laurareview,tye,flysoup}.

The idea that the selection of the correct vacuum should be driven by the quantum {\em dynamics} of gravity, is a tantalizing one. What we need is some physical requirement that can reduce the number of allowed initial states. Here we demonstrate that a superselection criterion emerges as a result of decoherence obtained through the backreaction of matter modes onto the gravitational degrees of freedom. The WMAP\cite{wmap} results might shed some insight \cite{steenlaura} into this. They are consistent with the predictions from inflation. In particular,  the anti-correlation between the TE and TT power spectra as a function of multipole number is a potential ``smoking-gun'' for inflation. Perhaps then, we should focus our attention on the portion of the landscape that allows for inflation.
 
 If inflation did indeed occur, the data argue for a high inflationary energy scale, perhaps near the grand unified scale $\sim 10^{16-17}$ GeV. How likely is this?  One way to frame this question is through the relation  $S_i ={3\pi G}\slash{\Lambda}$ between the entropy $S_i$ of the approximate de Sitter spacetime describing the inflationary phase, and the inflaton vacuum energy $\Lambda_i$. From this relation, we would infer that high scale inflation is extremely improbable, since the statistical probability $P$ is $\propto e^{S_i}$. 

It is important to note that all these arguments, which eventually lead to paradoxes and counterintuitive results, are statistical in nature and assume an equilibrium ensemble of initial inflationary patches. Many suggestions have been put forth \cite{entropy1,entropy2,albrecht} to resolve this problem,  but they all appear to lead to paradoxes when the state of the universe is evolved forward in time, especially if the endpoint of this evolution is a second DeSitter state in the far future. 

Our view is that the assumption of statistical equilibrium is not warranted in this setting. In particular, the fact that gravitational systems have negative specific heat makes equilibrium difficult, if not impossible to achieve. Dynamics must dictate whether high scale inflation will occur or not.  A more reasonable approach to the question of inflationary initial conditions would be a {\em dynamical} one. 

We exhibit such a dynamical mechanism in this work. The inclusion of the backreaction due to the quantum fluctuations of scalar perturbations gives rise to instabilities that render most of the inflationary patches {\em unstable} against gravitational collapse of super-horizon modes. This has the effect of dynamically reducing the allowed phase space of stable inflationary patches. This is essentially a Jeans instability effect, arising from the generation of tachyonic modes by the backreaction of the perturbations in Wheeler-deWitt (WdW) Master equation. We can then trace out the modes corresponding to collapsing patches to construct a reduced density matrix $\rho_{\rm red}$ for the patches that survive and enter an inflationary phase and use this to show explicitly that if $\hat{H}$ is the Hamiltonian of the system, $\left[\hat{H},\rho_{\rm red}\right]\neq 0$. This would imply that the initial states allowing for inflation do {\em not} form an equilibrium ensemble. 

This analysis can also be tied into current efforts\cite{tye,selection} to select appropriate vacua from the landscape of string vacua. We do this by taking the landscape as the configuration space for the wavefunction of the Universe so that the landscape minisuperspace can be thought of as the phase space of the initial conditions for the universe. In this construction, the minisuperspace of 3-geometries and string vacua is a real physical configuration space for the initial conditions, rather than an abstract metauniverse of unknown structure and unknown distribution of initial patches\cite{land1,land2}.

Our results have a number of implications that we will discuss in \cite{shortrichlaura}. In particular, to the question of the viability of casual patch physics, its implications for  holography, the $N$-bound proposal and Poincare recurrences.

Our plan of action will be first to represent as closely as possible what is known of the landscape vacua degrees of freedom distribution \cite{douglas,altland} and to construct wavepacket solutions of the WdW equation that correspond to classical trajectories of the universes on the landscape. Next, we will  perturb the landscape degrees of freedom along with the metric of 3-geometries and include their backreaction on the WdW equation. This will leads us to a Master equation for the probability distribution of vacua on the landscape. From this we will be able to infer the likelihood of inflationary initial conditions.

In the next section we discuss the model for the string landscape that we will use in the sequel. Sec.~\ref{sec:qcwdw} deals with the construction of the wavefunction of the Universe and its associated Wheeler-de Witt equation on our model of the landscape. After setting up the wavefunction, we turn to the issue of the backreaction of the massive modes in Sec.~\ref{subsec:backreact} and how they affect the evolution of the wavefunction. The dynamical selection mechanism is dealt with in Sec.~\ref{sec:select} and we conclude in Sec.~\ref{sec:conc}.


\section{A Model of the Stringy Landscape}
\label{sec:land}

As stated above, our goal is to investigate the dynamics of the wavefunction of the Universe defined on the string landscape. Given that we do not yet have as thorough a grasp on the structure of the landscape as we would like, we have to find a way to capture the features of the landscape that might be important for discussing inflationary initial conditions. 

We can do this in the following way.  In Ref.~\cite{land1,land2} the landscape was treated as a {\em disordered} lattice of vacua, where each of the $N$ sites is labelled by a mean value $\phi_i,\ i=1,\dots N$ of moduli fields. This allows us to use Random Matrix Theory (RMT) \cite{review1,review2,review3,efemetov} as well as other results from condensed matter systems. Each site has its own internal structure, consisting of closely spaced resonances around the central value. The disordering of the lattice is enforced via a stochastic distribution of mean ground state energy density $\epsilon_i,\ i=1\dots N$ of each site. These energies are drawn from the interval $\left[-W,+W\right]$, where $W\sim M_{\rm Planck}^4$ with a Gaussian distribution with width (disorder strength) $\Gamma$: $M_{\rm SUSY}^8\lesssim \Gamma\lesssim M_{\rm Planck}^8$, where $M_{\rm SUSY}$ is the SUSY breaking scale.

Quantum tunneling to other sites is always present which allows the wavefunction to spread from site to site. Together with the stochastic distribution of sites this ensures the Anderson localization\cite{anderson} of wavepackets around some vacuum site, at least for all the energy levels up to the disorder strength. This localization forces the wavefunction to remain within the non-SUSY sector of the landscape\cite{land1}. The energy density of the Anderson localized wavepacket is $\epsilon_i =|\Lambda_i +i\gamma|$, where $\Lambda_i$ is the vacuum energy density contribution to the site energy $\epsilon_i$ and $\gamma=l^{-1}l_p^{-3}$, where $l$ is the mean localization length and $l_p$ is the fundamental length of the lattice, which we will be take to be the Planck/string length. For large enough values of the disorder strength $\Gamma$, the majority of the levels are localized so that a semiclassical treatment of their classical trajectories in configuration space is justifed. 

To add gravity to this picture, we start by making use of the minisuperspace approximation, in which the scale factor $a$ of closed or flat $3$-geometries is added as a dynamical variable upon which the wavefunction will depend. In later sections, we will go beyond this approximation and add both metric and matter perturbations into the mix. 

As pointed out by Douglas et al \cite{douglas}, based on the two symmetries of this lattice, namely time-reversal invariance and rotation invariance, this sytem would fall in the same universality class as the CI-type class studied in Ref.\cite{altland} for quantum dots and random disordered systems.
However, in order to deal with realistic cosmologies, we include the scale factor $a$ of the 3-geometries as the gravitational degree of freedom, besides the landscape space of vacua which then breaks the time-invariance symmetry considered in Ref.~\cite{douglas}. The scale factor $a$ plays the role of an {\em intrinsic} time and the WdW equation for the wavefunction of the universe becomes manifestly {\em asymmetric} with respect to $a$. This has the implication that landscape plus gravity minisuperspace falls in the universality class of random lattices with broken time-invariance but with unbroken rotation-symmetry, which is the C-class of Ref.\cite{altland}.

In this picture, {\em each site} is a potential starting point for the universe since Anderson localization can occur in any of them. For this reason, the ensemble of sites, {\it i.e} the landscape minisuperspace is equivalent to the phase space of the initial conditions for the universe. 

Let us review some of the basic features of the RMT formalism, since this is one of the main tools we use to analyze the wave function of the Universe in our lattice picture of the landscape. The random matrix theory is achieved by taking many different realizations of the random potential of vacuum energies on the landscape. The mean averaged localization length $l$ of the wavefunction is obtained from the exponential decay of the retarded Green's function and given by the ensemble average of the norm of the retarded Green's function $G_R^{-1}$:
\be
\frac{l}{L}=\frac{1}{\pi}\langle1\left |\ln\left |\left |G^{-1}(\phi_i;\phi_j)\right|\right|\right |N\rangle \simeq \frac{2W <\epsilon_i>}{\Gamma},
\ee
where $L\simeq Nl_p$ is the size of the landscape sector.

The single-particle averaged density of states can be obtained from the imaginary part of the advanced Green's function $G_{A, jj}$, through the expression $\pi\ \rho(\epsilon)=\langle 1|{\rm Im} G_A|N\rangle$ . Note that $G_{A, jj}$ has poles at $|\epsilon_j|=|\Lambda_j -i\frac{\Gamma_j}{\epsilon_j}|$. Using RMT we can also write 
\begin{eqnarray}
\rho(\epsilon)&=& \frac{1}{N}\langle {\rm Tr} \delta(\epsilon - H(\phi))\rangle_{H_{\phi}}\nonumber \\
& =& \frac{1}{N\pi} \int{{\cal D}(\hat{H_{\phi}}) P(\hat{H_{\phi}}){\rm Im} (G_{A})}.
\end{eqnarray}

As discussed above, the non-SUSY sector of the landscape, with gravity switched on, belongs in the type C universality class. This allows us to write the joint probability distribution for the density of states as \cite{altland}:

\begin{equation}
P( \langle \hat{H}(\phi) \rangle)= P(\omega^2)\approx M_{\rm P}^{-8} \prod_{i\le j}({\omega_{i}^2}
 - {\omega_{j}^2})^2 \prod_k \omega_k^{2}\ e^{-\frac{\omega_k^2}{v^2}}.
\label{jointprobab}
\end{equation}
where $\epsilon_i = \omega_i^{2}$.
In the limit that the energy level spacing is less than $b=v\sqrt{M}$, where $M$ is the number of the internal degrees of freedom/sublevels in the $i$'th vacuum, (the closely spaced string resonances around the $i$'th vacua), this result goes to the familiar Wigner-Dyson result of random disordered systems, $P(\langle\hat{H}(\phi)\rangle =\omega^2) \approx \omega^2$.  We also see that for large energies, $P\approx (\omega^2 +\gamma - v^2)e^{-\omega^2(1/v^2 + l)}$.

\begin{figure}[!htbp]
\begin{center}
\raggedleft \centerline{\epsfxsize=3.5in \epsfbox{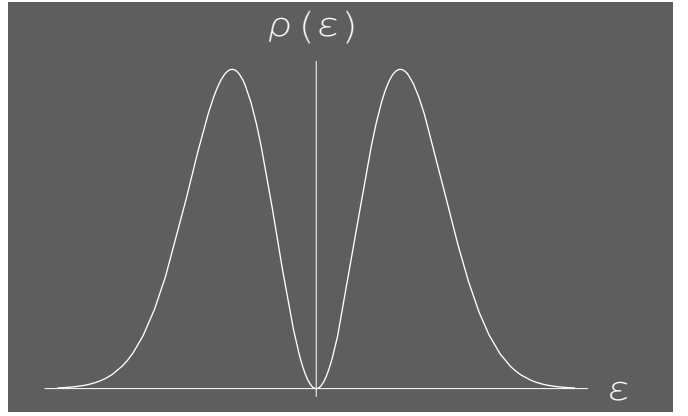}}
\caption{Density of states as a function of energy for the landscape system as derived from the Douglas and Altland probability distribution. Time-reversal invariance is preserved, {\it i.e.} gravity is not switched on yet.}
\label{fig:fig1}
\end{center}
\end{figure}

\begin{figure}[htbp]
\begin{center}
\raggedleft \centerline{\epsfxsize=3.5in \epsfbox{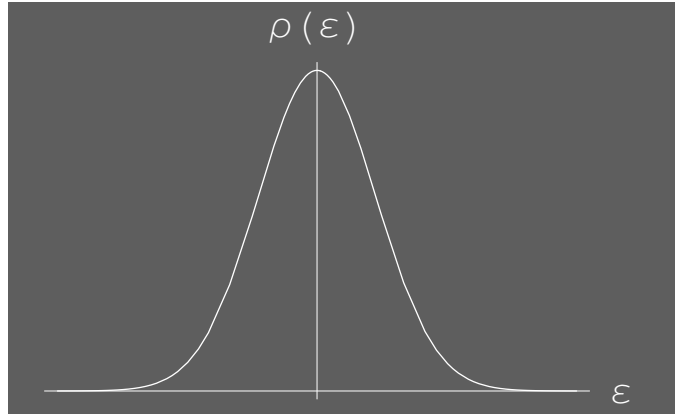}}
\caption{Density of states as a function of energy for the landscape +gravity system, namely when time invariance is broken. Notice the difference with the previous distribution around zero energy, namely  localized states exist at zero energy.}
\label{fig:fig2}
\end{center}
\end{figure}

The single-particle density of states $\rho(\omega)=\langle {\rm Tr} \delta(\omega^2 -H(\phi))\rangle$, obtained by integrating the above joint probability with respect to $\omega$, behaves as $\rho(\omega)\propto M_{\rm P}^{-8}  (1-\sin(l \omega^2)/l\omega^2)\ e^{-\frac{\omega^2}{v^2}}$ (Fig.\ref{fig:fig1}). When time reversal symmetry, given by the operation $\epsilon \rightarrow -\epsilon$, is broken, then $\rho(\omega)\simeq (1+\sin(l \omega^2)/l\omega^2)\ e^{-\frac{\omega^2}{v^2}}$ (Fig.\ref{fig:fig2}).

The 2-point correlation function for level-level mixing, $\langle\rho(\omega)\rho(\omega')\rangle$, which can be similarly obtained by the above averaging procedure with respect to the weight $P(\omega)$, goes to the Wigner-Dyson result for disordered systems,$\langle \rho(\omega)\rho(\omega')\rangle  \simeq  -\sin^2 (\pi l\omega^2)/(\pi l\omega^2)^2$.

These are all the results from RMT that we will need in the sequel. Naively, Fig.2 would imply that the most probable universe is the one residing around vacua with zero energy. This will change once the decoherence and backreaction effects of matter on the geometry are included.

\section{Quantum Cosmology and Wheeler-DeWitt Equation}
\label{sec:qcwdw}

It becomes clear that we need to study the quantum dynamics of gravity in combination with matter if want to address the issue of high energy inflation.
In order to make any progress, we restrict the number of degrees of freedom in the wave function of the Universe. This is usually done by invoking the minisuperspace approximation\cite{DeWitt}, where the wavefunction $\Psi$ depends on the scale factor $a(t)$, curvature $\kappa=0,\ \pm1$ of the FRW 3-geometries together with the landscape variables, collectively denoted by $\left\{\phi\right\}$ which will play the role of the massive modes in the Wheeler-DeWitt (WDW) equation.

The wave function of the universe $\Psi$ for the Friedman-Roberston-Walker (FRW) 3-geometries propagates on the landscape background with the vacuum distribution described in the previous section and parametrized by the collective coordinates $\{\phi\}= \{\phi_i^{n}\}$. Here $\phi_i$ is the central value of landscape variable on vacuum site $i=0,1,2,...N$ while $n$ counts the internal degrees of freedom within the $i$'th vacuum. The internal degrees of freedom are closely spaced resonances\footnote{ L.Susskind, private communication} around the mean value $\phi_i$, in the $i$'th vacuum site, within the energy range of the gaussian distribution of width $v$, $n=1,2..$. We can think of $n$ as counting the sublevels within the $i$'th energy level, and of the $\phi_1,...\phi_N$ as distinct energy levels.

Thus, our superspace consists of the infinite dimensional configuration space spanned by the variables $(a,\phi,f_n,d_n)$ where $f_n$ and $d_n $ denote the massive ($\phi$) and massless (metric) perturbation modes.

Before including the perturbation modes, $\{f_n,d_n\}$, the Wheeler-DeWitt equation for the wavefunction of the universe propagating on the minisuperspace spanned by the landscape variable $\phi$ and the FRW 3-geometries with line element
\begin{equation}
ds^2= -N^2 dt^2 + a^2(t) d{\bf x}^2 ,
\label{eq:lineelt}
\end{equation}  
is \cite{qcreview,wdw} 

\begin{eqnarray}
& &{\hat {\cal {H}}}\Psi(a,\phi) = 0 ~{\rm with} \nonumber \\
& &\hat{{\cal H}}=\frac{1}{2e^{3\alpha}}\left[\frac{4\pi}{3M_p^2}
\frac{\partial^2}{\partial\alpha^2}-
\frac{\partial^2}{\partial\phi^2}+V(\alpha, \phi)\right].
\label{eq:wdweq}
\end{eqnarray}
Here the scale factor $a$ has been written as $a=e^{\alpha}$ and $V(\alpha, \phi) = e^{6\alpha}m^2 \phi^2 - e^{4\alpha}\kappa, \kappa =0, 1$ for flat or closed universes.

If we change variables from $\phi$ to $x=e^{3\alpha}\phi$, we can rewrite the WdW equation in two separate equations in $x,\ \alpha$:
\bea
& &\Psi({\alpha, x})=\Sigma _j \psi_j(x)F_j(\alpha)\nonumber\\
& &\frac{3M_p^2}{4\pi} \left[
\frac{\partial^2}{\partial x^2}- V(x)\right]\ \psi_{j}(x) = \hat \epsilon_j \psi_{j}(x) \nonumber \\
& &-\frac{\partial^2}{\partial \alpha ^2}F_j(\alpha) = -\hat{\epsilon}_jF_j(\alpha),
\label{eq:sep}
\eea
where $\hat{\epsilon}_{j} = e^{6\alpha} \epsilon_j$.

The wavefunction $\Psi(\alpha,\phi)$ will in general be a superposition of many waves. In order to build wavepackets that correspond to classical paths in configuration space, some form of decoherence has to occur. Usually, this requires a separation between ``system'' and ``environmental'' variables; tracing over the environmental variables converts the ``system'' into an open one and allows it to behave classically. For our model of the landscape, we will take the higher super-horizon wavelength massive and metric multipoles $\{f_n,d_n\}$ to play the role of environmental variables. These modes couple with gravitational strength to the system $\Psi(\alpha,\phi)$. This coupling is of order $g\simeq{GM}\slash{R}$ with $M\approx O(M_{\rm Jeans}) \simeq H$ and $R\approx r_H\simeq H^{-1} $ so that $g\simeq H^2\slash M_P^2$. This is usually very small so that we can treat the higher multipoles as environmental variables and trace them out perturbatively~\cite{halliwellhawking,kiefer}.  

We now turn to the construction of wavepackets centered around a vacuum characterized by $\phi_i$~\cite{kieferwavepacket}. Using this wavepacket, we will then include the backreaction of the environment modes on this wavepacket. This will lead us from the WdW equation to a {\it Master Equation} for the wavefunction $\Psi(\alpha,\phi)$.
 
When we specialize these results to our version of the landscape, we consider the rescaled variables $ x=e^{3\alpha}\phi$, $\tilde{\omega}_k^{2}=e^{6\alpha}\ \omega_k^{2}$ and Eqns.\ref{eq:sep} lead to
\begin{eqnarray}
\hat{\cal{H}}(x)\psi_{j}(x) = \hat{\epsilon}_j \psi_{j}(x) ~{\rm where} \nonumber  \\
\hat{\cal{H}}(x)=\frac{3M_p^2}{4\pi} \left[
\frac{\partial^2}{\partial x^2} - \tilde{\omega}_k^{2}-\gamma)\right] \nonumber \\
\frac{\partial^2}{\partial \alpha ^2}F_j(\alpha) + (\tilde{\omega}_j^{2} - \gamma + \kappa e^{4\alpha})F_j&=&0 .
\label{eq:landwdweqn}
\end{eqnarray}

The localized solutions $\psi_j(x)$ around a vacuum site with energies centered around $\tilde{\omega_j}$ within the gaussian width $v$, are
 
 \be
 \psi_j (x)\simeq \sin(\tilde{\omega}_j x) \ e^{-\frac{(x-x')}{l} }.
 \ee 
 
The wavepacket is a superposition of these solutions for the $M$ internal degrees of freedom $n=1,...M$ with energies peaked around the mean value of site $x_j$, $\epsilon_j$, and amplitudes given by the Gaussian weight 
$$
A^{j}_{n} = \frac{l}{\pi\sqrt{Mv^2}}\ e^{- (\tilde{\omega}_n-\tilde{\omega}_0)^2 /(Mv^2)},
$$
namely\footnote{We will drop the index $j$ that labels the site from now on, keeping only the index $n$ that counts the internal degrees of freedom of the $j$'th  site}, 

$$
\psi(\tilde{\omega}_j)=\int_n d {\tilde{\omega}_n} A_n \psi_n F_n(\alpha).
$$
Within the WKB approximation, the turning points of the wavepacket are at  $\alpha=\alpha_n$ where $\alpha_n$ is a solution of: 

\begin{equation}
\tilde{\epsilon}_n^2 -\kappa e^{4\alpha_n}=0, \\
\label{eq:turnpoint}
\end{equation}
which in turn leads to the following solutions to the WdW equation:
\begin{equation}
F_j (\alpha)\approx \frac{1}{\sqrt{\tilde{\epsilon}_n (\alpha)}}
e^{-i \int_{\alpha_{n}}^{\alpha} \sqrt{\tilde{\epsilon}_n(\alpha^{\prime})}d\alpha^{\prime}}\
\label{eq:alpha}
\end{equation}

So, wave packets that are peaked around a level given by $\epsilon_n =\epsilon_0$ are constructed by the supersposition of the $M$ internal degrees of freedom of the landscape vacua, with the Gaussian weight $A_n$ given above (see Ref.~\cite{kieferwavepacket} for details of the construction).

For the sake of illustration we can consider closed universes with $\kappa=+1$. Then 

\begin{equation}
\Psi_j (\alpha,\phi) = \int_{-\infty}^{\infty} A_k F_k(\alpha)\psi_k (\alpha,\phi_j)d\tilde{\omega}_k
\label{eq:wavepacket}
\end{equation}

Eqn.~\ref{eq:wavepacket} shows that the inclusion of the internal degrees of freedom of vacuum $\phi_i$ results in a  Gaussian wavepacket spread with width $b^{-1}= (v\sqrt{M})^{-1}$. The wavepacket solution around some vacua $x_0$ of the $M$ internal oscillators with frequency levels $\tilde{\omega}^{2}_n =m_0a^3 (2n+1)$, where $n$ is a positive integer and $m_0 = V^{\prime \prime}_0$ is the curvature of the vacua potential, and  energies $\epsilon_n\sim|\tilde{\omega}^{2}_n\pm i\gamma|$, (where $\gamma =l^{-1}$) is

\bea
& &\Psi(x_0,\alpha)\simeq \int d\omega\ e^{-\frac{\omega^2}{b^2}}
\sin(\omega \ x_0)e^{-\gamma (x_0 -x)}e^{-i \alpha \omega}\nonumber \\
&=&\frac{b\sqrt{\pi}}{2 i}e^{-\gamma(x_0 -x)}\left(e^{-\frac{b^2}{4}(x_0+\alpha)^2}-e^{-\frac{b^2}{4}(x_0-\alpha)^2}\right).\nonumber\\
\eea

Tracing out the internal perturbation modes described by the index $M$ results in the reduced density  matrix $\rho_{\rm red}$ for the system $(\alpha,x)$ \cite{kiefer}:
\begin{equation}
\rho_{\rm red}(a, \phi;a^{\prime}, \phi^{\prime})\sim e^{-\frac{\Omega_{cl} M}{l}(a-a^{\prime})^2} \\
e^{-(b^2 - \frac{b^4}{4\gamma^2})^2 a^6 (\phi-\phi^{\prime})^2}.
\label{eq:density2}
\end{equation}
with $\Omega_{cl} = (m_{0}/M)^{1/2}$, $a=\exp\alpha$ and $b$ the width defined above.
We have ignored the contribution to the environment from the metric tensor perturbations $d_n$ in the above treatment since they are expected to be small compared to the massive modes (see \cite{halliwellhawking,kiefer}).

From the term depending on $(a-a^{\prime})^2$ in Eq.~\ref{eq:density2}, we see that the intrinsic time $a$ of the wavepacket  becomes classical first since the internal number of degrees of freedom $M$ is large, while $\phi$ becomes classical later when the scale factor grows larger than the Gaussian width.

The reduced density matrix above indicates how well the mean value $\epsilon_i,\phi_i$ can describe the vacuum site $i$ when the energy levels broaden due to the internal fluctuation modes of $\phi_i$. We expect the width $b=v\sqrt{M}$ to be at least of order SUSY breaking scale $M_{\rm susy}$, in order to account for the SUSY breaking of the zero energy levels $\omega_k=0$. Since the Fourier transform of the above wavepacket is still a Gaussian with width inverse that of $(x-x_0)^2$, we need  $b^2 < 2\gamma $ or $M_{\rm susy}\le M_*$, in order to have a meaningfully centered energy for the wavepacket made up from all the closed resonances (the internal degrees of freedom $M$). However, this gives rise to a spreading of the wavepacket in the moduli space $x$. To classicalize the system, we need to include the higher multipoles as environmental variables. We turn to this in the next section.

\subsection{Backreaction of Perturbations and the Master Equation}
\label{subsec:backreact}

The moduli fields as well as the metric have fluctuations about their mean value and those fluctuations can serve to decohere the wavefunction\cite{halliwellhawking}. This would then provide a classical probability distribution for scale factors and moduli fields. The procedure laid out in Ref.\cite{halliwellhawking} starts by writing the metric and the moduli fields as
\be
\label{eq:pert}
h_{ij} = a^2 \left(\Omega_{ij}+\epsilon_{ij}\right),\ \phi=\phi_0+\sum_n f_n(a) Q_n,
\ee
where $\Omega_{ij}$ is the FRW spatial metric, $\epsilon_{ij}$ is the metric perturbation (both scalar and tensor), $Q_n$ are the scalar field harmonics in the unperturbed metric and $f_n(a)$ are the massive mode perturbations. The index $n$ is an integer for closed spatial sections, and $k=n\slash a=n e^{-\alpha}$ denotes the physical wavenumber of the mode. As stated in Ref.\cite{halliwellhawking}, the fact that the CMB fluctuations are so small means that we can neglect the effects of the metric perturbations in the following calculations relative to the field fluctuations.

The wavefunction is now a function $\Psi=\Psi(a, \phi, \left\{f_n\right\})$. Inserting Eq.(\ref{eq:pert})   into the action, yields Hamiltonians $\left\{H_n\right\}$ for the fluctuation modes which, at quadratic order in the action, are decoupled from one another. The full quantized Hamiltonian $\hat{H} = \hat{H}_0 + \sum_n {\hat{H}}_n$ then acts on the wavefunction
\be
\label{eq:wavepert}
\Psi \sim \Psi_0 (a, \phi_0) \prod_n \psi_n (a, \phi, f_n).
\ee 
Doing all this yields the master equation
\be
\label{eq:master}
\hat{H}_0 \Psi_0 (a, \phi_0) = \left(-\sum_n \langle  \hat{H}_n\rangle\right) \Psi_0 (a, \phi_0),
\ee
where the angular brackets denote expectation values in the wavefunction $\psi_n$ and
\be
\label{eq:nham}
\hat H_n = -\frac{\partial^2}{\partial f_n^2} + e^{6 \alpha} \left( m^2 +e^{-2 \alpha} \left(n^2-1\right)\right) f_n^2,
\ee

\section{Dynamical Selection of Initial Conditions in the Phase Space of Inflationary Patches}
\label{sec:select}
Following Ref.\cite{kiefer2} a time parameter $t$ can be defined for WKB wavefunctions so that the equation for the perturbations $\psi_n$ can be written as a Schr$\ddot{\rm o}$dinger equation. 
If $S$ is the action for the mean values $\alpha, \phi$, define $y \equiv \left(\partial S\slash \partial \alpha\right)\slash \left(\partial S\slash \partial \phi\right)\sim \dot{\alpha}\slash \dot{\phi}$, so that we can write:
\bea
\label{eq:pertschr}
\psi_n &=& e^{\frac{\alpha}{2}} \exp\left(i \frac{3}{2 y} \frac{\partial S}{\partial \phi} f_n^2\right)\psi_n^{(0)}\nonumber \\
i\frac{\partial \psi_n^{(0)}}{\partial t}  &=& e^{-3 \alpha} \left\{-\frac{1}{2} \frac{\partial^2}{\partial f_n^2} + U(\alpha,\phi) f_n^2\right\} \psi_n^{(0)}\nonumber \\
U(\alpha,\phi) &=&e^{6\alpha} \left\{(\frac{n^2-1}{2})e^{-2\alpha} +\frac{m^2}{2} +\right .\nonumber \\
&+& \left . 9m^2 y^{-2}\phi^2
-6m^2 y^{-1}\phi\right\}.
\eea

During inflation, $S\approx-1\slash 3\ m e^{3\alpha} \phi_{\rm inf}$, where $\phi_{\rm inf}$ is the value of the field during inflation, so that $y=3\phi_{\rm inf}$. Thus long wavelength matter fluctuations are amplified during inflation and driven away from their ground state. After inflation, when the wavepacket is in an oscillatory regime, $y$ is large so that the potential $U(\alpha, \phi)$ changes from $U_{-}(\alpha, \phi)$ to $U_{+}(\alpha, \phi)$, where 
\be
\label{eq:posnegU}
U_{\pm} (\alpha, \phi) \sim e^{6\alpha} \left[\frac{n^2-1}{2}e^{-2\alpha}\pm \frac{m^2}{2}\right].
\ee
From Eq.(\ref{eq:pertschr}) we see that during inflation, the patches that have $U(\alpha, \phi)<0$, which can happen for small enough physical wave vector $k_n = n e^{-\alpha}$, develop tachyonic instabilities due to the growth of perturbations: $\psi_n \simeq e^{-\mu_n \alpha} e^{i \mu_n \phi}$, where $-\mu_n^2 = U(\alpha,\phi)f_n^2$. These trajectories in phase space {\em cannot } give rise to an inflationary universe, since they are damped in the intrinsic time $\alpha$ and so such modes do {\em not} contribute to the phase space of inflationary initial conditions. The damping of these wavefunctions is correlated with the tachyonic, Jeans-like instabilities of the corresponding mode $f_n$; when $U(\alpha, \phi)<0$, $f_n\sim e^{\pm \mu_n t}$, while for $U(\alpha, \phi)>0$, the long wavelength matter perturbations $f_n$ are frozen in. 

To see this more clearly, one can ask what happens to the massive perturbation $f_n$ modes in real spacetime for such damped wavefunction solutions. The equation of motion for $\phi,f_n$ can be obtained by varying the action with respect to these variables. For the tachyonic case $U<0$ universes, we have

\begin{equation}
\ddot{f}_n + 3H\dot{f}_n + \frac{U_{\pm}}{a_I^{3}} f_n =0,
\label{eq:modegrowth}
\end{equation}
where the inflation scale factor is $a_I =e^{3\alpha_I}$ and $U_{\pm}$ denotes the potential/(tachyonic) mass term case, Eqn.\ref{eq:posnegU}.

When $U<0$ one obtains growing and decaying solution in spacetime roughly  for $f_n \simeq e^{\pm \mu t}$. When $U>0$ then the $f_n$ are nearly frozen as in the standard perturbation theory case for superhubble wavelength modes.

This shows that, for damped universe solution in configuration space,with $U<0, \Psi\simeq e^{-\mu\alpha}$, the perturbation modes in real space $f_n$ grow rapidly. This corresponds to a universe that is collapsing instead of inflating due to the backreaction of massive super-Hubble perturbations $f_n$ which are coupled to the 3-geometry gravitationally via $U(\alpha,\phi)$. Note that the super-Hubble modes are {\it not} adiabatic and they do not re-enter in their ground state but rather in a highly excited state. For the inflating initial patches of our universe solutions, the superhorizon wavelength perturbations $f_n$ are nearly frozen, so we can ignore the energy corrections from the $\dot{f}_n$ terms. Notice that the cross-terms have also been dropped in the master equation since during inflation they are subleading compared to the quadratic terms included\cite{halliwellhawking,kiefer} with backreaction source term $\epsilon_n\simeq Ue^{6\alpha}$. 

What we glean from all this is that the following: all initial inflationary patches, characterized by values of the scale factor $a_{\rm inf}$ and Hubble parameter $h_{\rm inf}\equiv \sqrt{2\slash 3\pi} H_{\rm inf}\slash M_{\rm Planck} $ for which $U<0$ will collapse due to the backreaction of the superhorizon modes satisfying $k_n\leq m$. Since the backreaction effects due to modes with wavenumber $n$ scale as $a^{-2}$, patches for which $U>0$ will start to inflate and the backreaction effects will be inflated away. The surviving patches are then exactly those with 
\be
\label{eq:patchinf}
m^2 \phi_{\rm inf}^2 \simeq h_{\rm inf}^2 \geq k_n^2 = \left(\frac{n}{a}\right)^2\geq m^2\Rightarrow \phi_{\rm inf}\geq 1.
\ee
We have achieved our goal, namely we have shown that the quantum dynamics of the backreacting modes scours the Universe clean of regions which cannot support inflation! This reduction in the phase space of inflationary initial conditions implies that gravitational dynamics does {\em not} conserve the volume of the phase space, i.e. Liouville's theorem does not hold so that $\left[\hat{H}, \rho_{\rm red}\right]\neq 0$. 

The entropy can be obtained by taking the logarithm of the action above. However in order to simplify a rather messy expression for the action in our Master equation, let us take  the limit and think of the massive modes $f_n$ as collapsing into one black hole. Then we can write an  approximate expression for the entropy $S$ of the system of DeSitter patches together with the backreaction from the black hole ({\it i.e.} the massive modes), from our action including terms  up to quadratic order. This expression reduces to the entropy obtained by \cite{gibbonshawking} for Schwarschild-DeSitter geometries, with the identifications
\be
S \simeq (r_I -r_{f_{n}})^2,\  r_I \simeq H_I^{-1},\  r_{f_n} \simeq H_I^{-3/2}\langle \phi_I\sqrt{U}\rangle.
\label{desitterblackhole}
\ee
where $r_I$ denotes the De-Sitter horizon of the inflationary patches with Hubble parameter $H_I$ and $r_{f_N}$ the horizon of the ``black hole'' made up from the $f_n$, where $\langle f_n\rangle \simeq \phi_{\rm inf}$ and we have ignored numerical factors next to $r_I, r_{f_n}$ have been ignored.

It is interesting that the $U=0$ case, which can be thought of as a lower bound for the ``survivor'' patches, corresponds to the case of a zero entropy for the de Sitter-black hole system, {\it i.e.} when the surface gravity $r_I^{-1}$ of the de Sitter patch coincides with that of the black hole, $r_{ f_{n} }^{-1}$. This means that a black hole with the same horizon as the initial inflationary patch is the borderline between the damped and survivor universes, so that the zero entropy situation provides a lower bound on the initial conditions  $h_{\rm inf},\phi_{\rm inf}$ for an inflationary patch to appear and evolve into our universe. 

Note that in our model of the landscape as a stochastic lattice, the tracing out of the long wavelength fluctuations in the density matrix is encoded in the appearance of the mass scale, $\mu^2 = \langle U_n f_n^{2}\rangle$ and the internal dynamics of the wavepacket is encoded in the interplay between the SUSY scale and the landscape scale ($(l,b)$ or equivalently $(M_{SUSY},M_{*})$) in the reduced density matrix:
\bea
\rho &=& \int \Psi(\alpha,\phi,f_n)\Psi(\alpha',\phi',f'_n) \prod_n df_n df'_n \nonumber \\
&\simeq& \rho_0  e^{-\frac{(a\pi)^6 H^4 \mu^2(\phi-\phi')^2}{\phi_{\rm inf}^{2}} }\nonumber \\
\rho_0 &\simeq& \langle\Psi_0 (\alpha,\phi)\Psi_0(\alpha'\phi')\rangle \nonumber \\
&\simeq& e^{-M \Omega_{\rm cl} (a-a')^2}e^{-\Omega_{R} a^6 (\phi-\phi')^2}.
\label{eq:density1}
\eea
Here $2\Omega_R = b^2 - b^4 / 4\gamma$, $\Omega_{\rm cl} = \sqrt{m_0\slash M}$, where $m_0$ sets the scale for the frequency of the internal (resonance) oscillators and $M$ is the number of internal states we traced out initially. 
We have exhibited a lower bound on the energy scale for inflation in survivor universes, Eq.(4.4).
What happens if the initial fluctuation from the vacua minima $\phi_{\rm inf}$ is much larger than its lower bound? This is a difficult question to address since $\phi_{\rm inf} \gg \sqrt{\gamma}$ marks the breakdown of the semiclassical treatment. Nonetheless, we can extract some information by trying to extend our analysis to these cases. We've argued that for high scale inflation the backreaction of massive perturbations is negligible. Because of this, $[\hat{H},\rho]\approx 0$ and arguments based on Poincare recurrence phenomena may hold if quantum mechanics is valid in this regime.  But in this case, the Poincare recurrence time, $T_{\rm recurr}\simeq e^{S}$ is short so that these patches become quantum on times scales of order $T_{\rm recurr}$. Demanding that recurrence time is as large as the age of the Universe, or equivalently that the broadening of this energy level $\delta E = e^{-S}$ should be less the difference between energy levels $\delta E \ll \gamma$ provides an upper bound on the field values at which quantum entanglement occurs over long enough times such that it allows inflation to start.  We conclude that for $b^2<\Lambda_{\rm inf}< \gamma$, the backreaction of the superhorizon modes included can be roughly approximated by Eq.(\ref{desitterblackhole}).

\section{Discussion and Conclusions}
\label{sec:conc}

Why did the Universe start in a state of lower than anticipated entropy? Equivalently, how did high scale inflation occur? The key to answering these questions is to not be fooled by arguments based on {\em equilibrium} statisitical mechanics. In fact, it is exactly the {\em non}-equilibrium dynamics of superhorizon modes and their backreaction onto the mean values of  $a,\ \phi$ that selects out the regions which inflate; patches that do not satisfy $m<H_I, l<\phi_I<b$ will recollapse. This non-equilibrium dynamics also leads to non-ergodic behavior in the phase space of initial conditions, as well as entanglement of states. This last is significant, since it implies that a holographic description of gravity during inflation may not be tenable. Our analysis also gives rise to questions about the applicability of the causal patch and $N$-bound approaches to inflation that we discuss in \cite{land1}.

Despite having made use of a particular model of the landscape to arrive at our results we would argue that our results should have wider applicability. The landscape minisuperspace serves mostly to provide a concrete realization of our approach, specifically the scales $M_*, M_{\rm SUSY}$ for the widths of the initial inflationary patches. The rest of the quantum cosmological calculation based on backreaction and the master equation is general and could be applied to any phase space for the initial conditions once its structure was known. What we have learned here is that any model of a universe containing both matter and gravity will exhibit this non-ergodic behavior driven by out-of-equilibrium dynamics. In fact, such universes will experience a {\it superselection rule for the Initial Conditions}. Since non-ergodicity compresses the volume $V$ of phase space available to survivor universes, thereby lowering the entropy $S \simeq \log V$ of survivor universes, the low entropy from the reduction of phase space, for the survivor initial patches provides an explanation for the observed arrow of time in high scale inflation. 

Is our model predictive?  In a forthcoming paper\cite{richlauranext} we will report how remnants of quantum entaglement between in and out modes, as represented by the cross-terms in the reduced density matrix, might be tested by cosmological observables such as nongaussianities in CMB and large scale structure.

\begin{acknowledgments}
R.H. was supported in part by DOE grant DE-FG03-91-ER40682.
LMH was suported in part by DOE grant DE-FG02-06ER41418 and NSF grant PHY-0553312.
\end{acknowledgments}


\end{document}